\documentclass[twoside]{article}
\usepackage{amssymb}

\newtheorem{defi}{Definition}[section]
\newtheorem{prop}[defi]{Proposition}

\newcommand{\ket}[1]{|#1\rangle}
\newcommand{\mycomment}[1]{}

\begin{document}
\setlength{\textheight}{8.0truein}    %

\normalsize
\thispagestyle{empty}
\setcounter{page}{1}

\vspace*{0.88truein}

\centerline{\bf Quantum Advantage without Entanglement}
\vspace*{0.37truein}
\centerline{\footnotesize
Dan Kenigsberg, Tal Mor and Gil Ratsaby}
\vspace*{0.015truein}
\centerline{\footnotesize\it 
 Computer Science Department, Technion, Haifa 32000, Israel.}
\centerline{\footnotesize \{danken,talmo,rgil\}@cs.technion.ac.il}
\vspace*{0.21truein}

\abstract{
We study the advantage of pure-state quantum computation without
entanglement over classical computation. For the Deutsch-Jozsa
algorithm we present the \emph{maximal} subproblem that can be
solved without entanglement, and show that the algorithm still has
an advantage over the classical ones. We further show that this
subproblem is of greater significance, by proving that it contains
\emph{all} the Boolean functions whose quantum phase-oracle is
non-entangling. For Simon's and Grover's algorithms we provide
simple proofs that no non-trivial subproblems can be solved by
these algorithms without entanglement.
}

\section{Introduction}

The fusion of quantum theory and computer science has introduced
new capabilities to the world of computation and communication.
Utilizing these capabilities led to some spectacular results:
Shor's factorization algorithm~\cite{Shor97}, Grover's quantum
search algorithm~\cite{Grover96}, Bennett-Brassard's quantum key
distribution~\cite{BB84} and quantum
teleportation~\cite{Teleport}, all extend beyond well-believed
bounds.

The source of quantum computation power is still debated. Some
argue that this power arises because unlike classical systems, the
state of a multipartite quantum system cannot always be considered
as a mere correlated combination of its subsystems. Such states
are said to be \emph{entangled}, as opposed to \emph{separable}
states. Entanglement appears in most of the great achievements of
quantum information theory, from Shor's factorization
algorithm~\cite{Shor97} and other
algorithms~\cite{Grover96,DJ92,Simon94} to quantum
teleportation~\cite{Teleport}, superdense
coding~\cite{Superdense}, quantum error correction~\cite{Shor95},
and some quantum key distribution
schemes~\cite{Ekert91,BBM92,BHM96}.
It is uncontested that entanglement is a key resource of quantum
computation, or in the words of Micha\l{}
Horodecki~\cite{MHorodecki01}, entanglement is ``the corner-stone
of the quantum information theory''. When no entanglement exists
in a pure-state quantum algorithm, the computation can be
simulated efficiently and exactly using classical
means~\cite{Jozsa98book}. Furthermore, when entanglement exists but
its amount is bounded, Jozsa and Linden showed~\cite{JL02} that
the computation can still be efficiently simulated classically by
a coin-tossing algorithm. However, their work does not rule out
significant advantage of quantum computation without entanglement
(\textbf{QCWE}) in an oracle-based setting (e.g. for the
Deutsch-Jozsa algorithm). It also does not rule out exponential
advantage of mixed-state QCWE over probabilistic classical
computation.

Some cases have been found where even without entanglement,
quantum computation outperforms classical computation. Collins,
Kim and Holton~\cite{Collins98} solve the
Deutsch-Jozsa~(DJ)~\cite{DJ92} problem without entanglement, but
only for $n=2$ bits, and prove that entanglement is required for
any larger $n$; Braunstein and Pati~\cite{BraunsteinPati02} show
that using pseudo-pure states, Grover's search problem can be
solved without entanglement for $n\leq3$ bits more efficiently
than classically; Lloyd~\cite{Lloyd00}
suggests an entanglement-free implementation of Grover's
algorithm, but with exponential spatial complexity; Biham,
Brassard, Kenigsberg and Mor~\cite{BBKM02} use a non-standard
computation model, with a limitation on the number of allowed
queries, to prove a tiny separation for any $n$ in the context of
Deutsch-Jozsa's and Simon's~\cite{Simon94} problems (with mixed
states); Meyer~\cite{Meyer00soph} notes that Bernstein-Vazirani
algorithm~\cite{BV97,TerhalSmolin98} requires no entanglement, yet
uses only one oracle call, while the best classical algorithm
requires $n$ oracle calls.

In this paper we investigate the advantage of pure-state QCWE in
several quantum algorithms. We introduce a restricted version of
the Deutsch-Jozsa problem for which the algorithm generates
\emph{no entanglement}\footnote{A restricted problem is referred
to as a \emph{subproblem}, i.e., a subset of the legal inputs. In
this case, the subproblem is a subset of the legal oracles.}. We
show that the algorithm still has a significant advantage over the
corresponding classical complexity, and prove that this restricted
problem is \textbf{maximal}, in the sense that any extension of it
will generate entanglement. This gives some evidence that for the
Deutsch-Jozsa problem there is no `real' exponential gap between
QCWE and exact classical algorithms\footnote{However, see open
questions in Section~\ref{sec:summary}.}. Furthermore, we show the
significance of this subproblem, and prove that it contains
\emph{any} separability-conserving phase-oracle. We then move on
to Simon's problem and Grover's algorithm, and show that no
non-trivial instance of these problems can be solved without
entanglement.

We use the following conventions: the \textit{quantum oracle} of a
Boolean function $f:\{0,1\}^{n}\to\{0,1\}$, is the black-box
unitary operation $U_f$, which for an $(n+1)$-qubit input state
$\ket{x}\ket{y}$, outputs $\ket{x}\ket{y\oplus f(x)}$. The
\textit{quantum phase-oracle} of $f$, is the black-box unitary
operation $V_f$, which for an n-qubit input state $\ket{x}$,
outputs $(-1)^{f(x)}\ket{x}$. Note that when $\ket{y}=\ket{-}$,
$U_f$ functions as a phase-oracle.

\section{Entanglement in the Deutsch-Jozsa Algorithm}

In this section we analyze the occurrence of entanglement in the
execution of the Deutsch-Jozsa Algorithm. We present a restricted
version of the problem, which is entanglement-free, maximal, and
yet advantageous over the best exact classical algorithm.

\subsection{The Deutsch-Jozsa Algorithm}

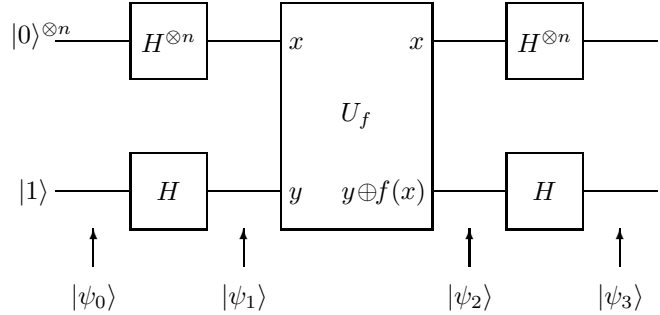
\begin{figure}
\setlength{\unitlength}{1cm}
\begin{center}
\begin{picture}(10,5)
\put(0.5,1.9){$\ket1$} \put(0.4,3.97){$\ket0^{\otimes n}$}

\put(1,2){\line(1,0){1}} \put(2,1.5){\framebox(1,1){$H$}}
\put(3,2){\line(1,0){1}}

\put(1,4){\line(1,0){1}} \put(2,3.5){\framebox(1,1){$H^{\otimes
n}$}} \put(3,4){\line(1,0){1}}

\put(4,1.5){\framebox(2,3){$U_f$}} \put(4.1,1.9){$y$}
\put(4.8,1.9){$y\!\oplus\!\!f(x)$} \put(4.1,3.9){$x$}
\put(5.7,3.9){$x$}

\put(6,2){\line(1,0){1}} \put(7,1.5){\framebox(1,1){$H$}}
\put(8,2){\line(1,0){1}}

\put(6,4){\line(1,0){1}} \put(7,3.5){\framebox(1,1){$H^{\otimes
n}$}} \put(8,4){\line(1,0){1}}

\put(1.5,1){\vector(0,1){0.5}} \put(1.2,0.5){$\ket{\psi_0}$}
\put(3.5,1){\vector(0,1){0.5}} \put(3.2,0.5){$\ket{\psi_1}$}
\put(6.5,1){\vector(0,1){0.5}} \put(6.2,0.5){$\ket{\psi_2}$}
\put(8.5,1){\vector(0,1){0.5}} \put(8.2,0.5){$\ket{\psi_3}$}

\end{picture}
\caption{The Deutsch-Jozsa algorithm (which is a quantum
subroutine common to Simon, Grover, Bernstein-Vazirani and
other algorithms)}
\label{fig:DJ}
\end{center}
\end{figure}

Let $f$ be a Boolean function $f:\{0,1\}^{n}\to\{0,1\}$, with a
promise that $f$ is either \emph{constant} or \emph{balanced},
namely, the value of $f$ is either the same for all the members in
its domain, or it is 1 for exactly half of it and 0 for the other
half. The function $f$ is given as an oracle, and our goal is to
discover whether it is constant or balanced. Note that a
deterministic classical algorithm that solves this problem must
perform $2^{n}/2+1$ oracle queries. The Deutsch-Jozsa
algorithm~\cite{DJ92}, represented by the quantum circuit in
Figure~\ref{fig:DJ}, distinguishes between the two possible types
of $f$ using only one quantum query.
\newline\newline
The quantum register is changed by the algorithm steps as follows:

\begin{enumerate}
\item The initial state is $\vert\psi_{0}\rangle=\vert0\rangle^{\otimes n}\vert1\rangle$.
\item After applying $n+1$ Hadamard gates:
    \begin{displaymath}
        \vert\psi_{1}\rangle=  \sum_{x\in\{0,1\}^n}
            \frac{\vert x \rangle}{\sqrt{2^n}}
            \big[ \frac{\vert 0 \rangle - \vert 1 \rangle} {\sqrt{2}} \big].
    \end{displaymath}
\item Applying the $f$ quantum query yields:
    \begin{displaymath}
        \vert\psi_{2}\rangle=  \sum_{x\in\{0,1\}^n}
            \frac{(-1)^{f(x)}\vert x \rangle}{\sqrt{2^n}}
            \big[ \frac{\vert 0 \rangle - \vert 1 \rangle} {\sqrt{2}} \big].
    \end{displaymath}
\item Finally, after applying the last $n$ Hadamard gates:    \begin{displaymath}
        \vert\psi_{3}\rangle=  \sum_{z\in\{0,1\}^n}
            \sum_{x\in\{0,1\}^n}
            \frac{(-1)^{x\cdot z+f(x)}\vert z \rangle}{2^n}
            \big[ \frac{\vert 0 \rangle - \vert 1 \rangle} {\sqrt{2}} \big].
    \end{displaymath}
\end{enumerate}
The amplitude of $\ket{z} = \ket{0}^{\otimes n}$ in
$\ket{\psi_{3}}$ is $\sum_{x}(-1)^{f(x)}/2^{n}$. Thus, if $f$ is
constant we obtain this state with certainty when measuring
$\vert\psi_{3}\rangle$, and if $f$ is balanced it is certain that
some other state is measured. It follows that after just one query
to the oracle, we are able to determine with certainty whether $f$
is constant or balanced.

\subsection{The Deutsch-Jozsa Algorithm without Entanglement}

Bernstein and Vazirani defined a promise problem~\cite{BV97} that
can be solved by the Deutsch-Jozsa subroutine using a single
oracle call. Classically, this problem requires $n$ oracle calls.
Later, Meyer~\cite{Meyer00soph} noted that entanglement is not
generated during the execution of the algorithm on their problem.
Since the BV promise set is a subset of the DJ promise set, a
corollary from these results is that a subclass of the DJ problem
requires no entanglement.

We follow a different route, aiming to find the maximal
entanglement-free set. Looking at the Deutsch-Jozsa algorithm, we
note that the only step in which entanglement can be generated is
the third step, where the $f$ oracle is applied to the quantum
register. We define the following restricted DJ promise problem:
the function $f$ is again either balanced or constant, but it is
also promised to be \emph{\textbf{non-entangling}} for the DJ
algorithm, i.e., applying the oracle to $\ket{\psi_{1}}$ yields a
separable state (the state $\ket{\psi_{2}}$). In order to solve
this promise problem we execute the original algorithm. Since this
is a sub-problem of the original one, the algorithm's correctness
is assured and one quantum query is enough to determine whether
the function is balanced or constant. Note also that if there
exist such non-entangling balanced functions (i.e., the constant
functions are not the only possible non-entangling functions),
classical algorithms cannot solve the problem with only one query,
and so are inferior compared with the quantum one.\newline

\begin{defi}
Let $F_{DJ}^{\otimes}$ be the set of all Boolean functions
$f:\{0,1\}^{n}\to\{0,1\}$ of the following form:
    \begin{equation}
    f(x) = (a \cdot x)\oplus c,
    \label{eq:nonent}
    \end{equation}
where $a \in \{0,1\}^{n}$, $c \in \{0,1\}$ and `$\cdot$' is the
inner product modulo 2. For a given $c$ and $a$, we denote the
corresponding $f$ by $f_{c,a}$.
\end{defi}

\begin{prop}
Any $f_{c,a} \in F_{DJ}^{\otimes}$ is either constant or
balanced.
\end{prop}
\textbf{Proof.} First note that if $a=00 \cdots 0$, $f_{c,a}$ is constant.
For non-zero $a$ the function $f_{a,c}(x)$ is balanced. This is clear since
the inhomogeneous linear equation system
\[a\cdot x\oplus c=0\]
implies one linear constraint over an $n$-dimensional $x$, and
therefore the solution space is $(n-1)$-dimensional. This means that out
of $2^n$ possible $x$'s, $2^{n-1}$ (half) are solutions.
\hfill$\square$

Denote by $DJ^{\otimes}$ the
``entanglement-free"\footnote{Proposition~\ref{prop:DJ_sep}
provides the justification for this name.} Deutsch-Jozsa problem,
defined for the set $F_{DJ}^{\otimes}$ instead for any
constant/balanced function.
\begin{prop}\label{prop:DJ_sep}
Entanglement is never generated when executing the Deutsch-Jozsa
algorithm for $DJ^{\otimes}$.
\end{prop}
\textbf{Proof.} For an  input $x=x_1x_2 \cdots x_n$, we
denote by $J_x$ the support of $x$, i.e., the set of indexes
of nonzero elements in $x$.
Applying the oracle $f_{c,a}$ on $\vert\psi_{1}\rangle$
yields:
    \begin{eqnarray}
        \vert\psi_{2}\rangle &=&
        \sum_{z\in\{0,1\}^n}
            \frac{(-1)^{f_{c,a}(z)}\vert z \rangle}{\sqrt{2^n}}
            \big[ \frac{\vert 0 \rangle - \vert 1 \rangle} {\sqrt{2}} \big] \nonumber \\
        &=&
         \sum_{z\in\{0,1\}^n}
            \frac{(-1)^c(-1)^{z \cdot a} \vert z \rangle}{\sqrt{2^n}}
            \big[ \frac{\vert 0 \rangle - \vert 1 \rangle} {\sqrt{2}} \big] \nonumber \\
        &=&
         \frac{(-1)^c}{\sqrt{2^n}}
         \sum_{z\in\{0,1\}^n}
            (-1)^{\sum_{J_z} a_i} \vert z \rangle
            \big[ \frac{\vert 0 \rangle - \vert 1 \rangle} {\sqrt{2}} \big] \nonumber \\
        &=&
        \frac{ (-1)^c}{\sqrt{2^n}}
        (\vert0\rangle+(-1)^{a_{1}}\vert1\rangle)\cdots(\vert0\rangle +(-1)^{a_{n}}\vert1\rangle)
            \big[ \frac{\vert 0 \rangle - \vert 1 \rangle} {\sqrt{2}}
            \big], \label{eq:star}
    \end{eqnarray}
thus for any $f_{c,a}$, $\vert\psi_{2}\rangle$ is not entangled.
\hfill$\square$

Having established these properties of $DJ^{\otimes}$, we would
like to show that it is maximal, in the sense that it includes any
case of separable computation for the Deutsch-Jozsa algorithm.

\begin{prop}
Any non-entangling balanced function $f$ equals $f_{c,a}$ for
some $c \in \{0,1\}$ and $a \in \{0,1\}^n$.
\end{prop}
\textbf{Proof.} Let $\vert\psi_{2}\rangle$ be as defined for the
Deutsch-Jozsa algorithm. If $f$ is non-entangling, then:
    \begin{eqnarray}
        \vert\psi_{2}\rangle &=&
        \sum_{x\in\{0,1\}^n}
            \frac{(-1)^{f(x)}\vert x \rangle}{\sqrt{2^n}}
            \big[ \frac{\vert 0 \rangle - \vert 1 \rangle} {\sqrt{2}}\big] \label{eq:psi2_a}\\
        &=& e^{i\varphi_0}
        \bigotimes_{k=1}^{n}(\cos\theta_k\vert 0 \rangle+e^{i\varphi_k}\sin\theta_k\vert 1 \rangle)
        \big[ \frac{\vert 0 \rangle - \vert 1 \rangle} {\sqrt{2}}\big]. \label{eq:psi2_b}
    \end{eqnarray}
We compare the coefficients of Eqs.~(\ref{eq:psi2_a})
and~(\ref{eq:psi2_b}). First note that $\varphi_0$ must be 0 or
$\pi$ since the phase of the state $\vert00\cdots 0\rangle$ is
$\pm$1. Suppose that for some $1\le k \le n$, $|\cos\theta_k| \ne
|\sin\theta_k|$. Then the coefficients of the states
$\vert00\cdots 000 \cdots 0\rangle$ and $\vert00\cdots 010 \cdots
0\rangle$ where the 1 is on the $k$'th position, are not of the
same magnitude, in contradiction. Similarly, if for some $1\le k
\le n$, $\varphi_k \notin \{0,\pi \}$, then the state
$\vert00\cdots 010 \cdots 0\rangle $ where the 1 is on the $k$'th
position, has a complex coefficient, in contradiction. It follows
that $\vert\psi_{2}\rangle$ is a product state of exactly the form
of Eq.~(\ref{eq:star}), which corresponds to a function $f_{c,a}$
as claimed.
\hfill$\square$

We now show that solving $DJ^{\otimes}$ on an exact classical
computer is not a trivial task.
\begin{prop}
The classical computational complexity of $DJ^{\otimes}$ is
$\Theta (n)$.
\end{prop}
\textbf{Proof.}
For a function of the form $f(x) = (a \cdot x)\oplus c$, one must know all the bits of $a$ in order to
check whether $f$ is constant or balanced. Even one missing bit can determine the function's type either way.
An oracle query of $f$ yields an equation of the form $(a \cdot x)\oplus c = b$ for $b \in \{0,1\}$,
which is a linear Boolean equation in $n$ variables: the $n$ bits composing $a$. Therefore, in order to find
$a$ out, one must consider at least $n$ equations. This
means that any classical algorithm will require at least $n$
evaluations of $f$ in order to exactly find $a$. Note also
that $n$ evaluations are enough, since the evaluation of $f(2^k)$ yields the $k$th bit of $a$,
so evaluating $f$ on the $n$ powers of 2 will determine it uniquely.
\hfill$\square$

Note that much like $DJ$, $DJ^{\otimes}$ can be solved with a
constant number of oracle queries if errors are permitted.

\subsection{General Phase Oracles and $DJ^{\otimes}$}

We now show that the $DJ^{\otimes}$ problem is related to the
whole set of separability-conserving quantum phase-oracles.

\begin{defi}
An operation $U$ is separability-conserving, if for any separable
state $\ket{\psi}$, $U\ket{\psi}$ is separable.
\end{defi}

\begin{prop}
Let $f:\{0,1\}^n \rightarrow \{0,1\}$ be a Boolean function and
$V_f$ the corresponding quantum phase-oracle. If $V_f$ 
is separability-conserving,
then $f \in
F_{DJ}^{\otimes}$. \label{sep-conserve}
\end{prop}
\textbf{Proof.}
A separable state $\ket{\Psi}$ can be written as
\begin{eqnarray}
\ket{\Psi} &=& e^{i\varphi_0}
        \bigotimes_{k=1}^{n}(\cos\theta_k\vert 0 \rangle+e^{i\varphi_k}\sin\theta_k\vert 1 \rangle) \label{eq:sep1}\\
        &=& \sum_{x=0}^{2^n-1} \alpha_x \ket{x} \label{eq:sep2}
\end{eqnarray}
for some $0 \leq \varphi_k,\theta_k \leq \pi$ and $\alpha_x \in
C$. Note that when $\cos\theta_k=0$ or $\sin\theta_k=0$, we choose
$\varphi_k=0$ without loss of generality. Since $V_f\ket{\Psi}$ is
separable too, it holds that:
\begin{eqnarray}
V_f\ket{\Psi} &=&\sum_{x=0}^{2^n-1} \tilde\alpha_x \ket{x}=
        \sum_{x=0}^{2^n-1} (-1)^{f(x)}\alpha_x \ket{x} \label{eq:ufsep1} \\
        &=&
        e^{i\tilde{\varphi_0}}
        \bigotimes_{k=1}^{n}(\cos\tilde{\theta_k} \vert 0 \rangle+e^{i\tilde{\varphi_k}}\sin\tilde{\theta_k}\vert 1 \rangle) \label{eq:ufsep2}
\end{eqnarray}
for some $0 \leq \tilde{\varphi_k},\tilde{\theta_k} \leq \pi$. Let
$e_k$ denote the binary string with $1$ in the $k$th bit and $0$
in all the rest. First observe that $\alpha_x = \pm
\tilde\alpha_x$ for all $x$ and that there must be at least one
$x$ such that $\alpha_x\neq0$. For any $y$, including $y=x\oplus
e_k$, $\alpha_y= \pm \tilde\alpha_y$, we may write
\[\frac{\alpha_y}{\alpha_x}= \pm \frac{\tilde\alpha_y}{\tilde\alpha_x}\]
which means that $|\tan\theta_k|=|\tan\tilde\theta_k|$
 or in other words:
\begin{equation}
\cos\theta_k = \pm\cos\tilde{\theta_k},~
\sin\theta_k=\pm\sin\tilde{\theta_k}.
\end{equation}
Let $J=\{k:\cos\theta_k=0\}$ be the set of qubits whose
$\cos\theta_k$ is zero, and choose a string $y$ so that $y_k=1
\Leftrightarrow k\in J$. Note that
\[\tilde\alpha_y=\pm e^{i\tilde\varphi_0}\prod_{k\notin J}\cos\tilde\theta_k=
(-1)^{f(y)} e^{i\varphi_0}\prod_{k\notin J}\cos\theta_k=\alpha_y
(-1)^{f(y)}\neq0.\] From $\prod_{k\notin J}\cos\theta_k = \pm
\prod_{k\notin J}\cos\tilde{\theta_k}\neq0$ we obtain
$e^{i\tilde{\varphi_0}} = \pm e^{i\varphi_0}$. For $j\notin J$,
looking at $\tilde\alpha_{y\oplus e_j}$:
    \begin{eqnarray*}
        \tilde\alpha_{y\oplus e_j}
        &=&
            e^{i\tilde\varphi_0}e^{i\tilde\varphi_j}\sin\tilde\theta_j
            \prod_{k\notin J,k\neq j}\cos\tilde\theta_k\\
        &=& (-1)^{f(x)}e^{i\varphi_0}e^{i\varphi_j}\sin\theta_j \prod_{k\notin J,k\neq j}\cos\theta_k
         = \alpha_{y\oplus e_j} (-1)^{f(x)}
    \end{eqnarray*}
similarly leads to $e^{i\tilde{\varphi_j}} = \pm e^{i\varphi_j}$.
For $j\in J$ we have $\cos\theta_j=\cos\tilde\theta_j=0$ and by
definition $\varphi_j=0=\tilde{\varphi_j}$.

Having established that $\cos\theta_k = \pm\cos\tilde{\theta_k}$,
$\sin\theta_k=\pm\sin\tilde{\theta_k}$ and $e^{i\tilde{\varphi_k}}
= \pm e^{i\varphi_k}$ for any $k$, it follows that there exist
$c,a_k\in\{0,1\}$ such that
\[
V_f\ket{\Psi} =
    e^{i\varphi_0}
        (-1)^c\bigotimes_{k=1}^{n}(\cos\theta_k\vert 0 \rangle+(-1)^{a_k} e^{i\varphi_k}\sin\theta_k\vert 1 \rangle)
\]
which is exactly $V_{f_{c,a}}\ket{\Psi}$.

Now, it is enough to look at the separable state $\ket{\Psi_H} =
\frac{1}{\sqrt{2^n}}\sum_{x\in\{0,1\}^n}\ket{x}$. Since the above
result applies to any separable state, it also applies to
$\ket{\Psi_H}$, therefore, there exist $c$ and $a$ such that
$V_f\ket{\Psi_H} = V_{f_{c,a}}\ket{\Psi_H}$. We note that the
$x$th coefficient of $V_f\ket{\Psi_H}$ determines $f(x)$ to be
$(-1)^{f_{c,a}(x)}$ since this is exactly the phase added by
$f_{c,a}$ to the state $\ket{x}$. Since this is true for all the
$2^n$ coefficients, it follows that $f \equiv f_{c,a}$, thus $f
\in F_{DJ}^{\otimes}$.
\hfill$\square$

In conclusion, we identified the \emph{maximal} subset of the
Deutsch-Jozsa problem, which is solved with one quantum query by
the Deutsch-Jozsa algorithm \emph{without entanglement}, while the
best exact classical algorithm requires a linear number of calls.
We showed that the significance of this subset reaches beyond the
scope of DJ problem: \emph{any} non-entangling phase oracle is an
oracle of a function from this subset. Note also that here the
quantum-to-classical gap in the exact case diminishes from
$O(2^n)$:$O(1)$ to $O(n)$:$O(1)$. This is the price we pay for not
using entanglement.

We remark that it can now be seen that the function set
of~\cite{BV97,Meyer00soph} contains \emph{exactly} half of the
possible non-entangling functions.

\subsection{Separable Implementation of the Oracle}

It may be claimed that even though there is no entanglement after
any step in the algorithm, the oracle must use entanglement during
the computation of the function. We show here that there is an
entanglement-free implementation for the oracle. Consider the
following transformation:

\begin{displaymath}
\vert x \rangle \rightarrow (-1)^{c+1} \bigotimes_{i=1}^n
(-1)^{a_{i} \cdot {x_i}} \vert x_i \rangle
\end{displaymath}
It is easy to see that the tensor product on the right-hand side
gives exactly the requested result of applying $f_{c,a}$ on
input $\vert x \rangle$, i.e., $(-1)^{f_{c,a}} \vert x
\rangle$. The oracle operation can be done locally, using $n$ single-qubit
transformations such as
\[\left(\begin{array}{cc}1&0\\0&e^{ia_i\xi_i(t)}\end{array}\right),\]
where $\xi(t)$ ascends from $0$ to $2\pi$,
maintaining separability even in continuous time.

\section{Simon's Problem}

Simon presented~\cite{Simon94} the following oracle problem:\\
Let $f:\{0,1\}^n \rightarrow \{0,1\}^n$ be a 2-to-1
function\footnote{In some versions it is also allowed to be
constant.}, such that
\begin{displaymath}
\forall x \neq y:~ f(x)=f(y) \Leftrightarrow y=x\oplus a,
\end{displaymath}
where $a$ is a fixed $n$-bit string called the function's period, and $\oplus$ is the bitwise XOR operation.
The goal is to determine the value of $a$.

In order to solve this problem classically with high probability,
the oracle must be queried an exponential number of times.
However, the following quantum procedure solves it with high
probability using a polynomial number of queries.

\begin{enumerate}
\item The initial state is $\vert\psi_{0}\rangle=\vert0\rangle^{\otimes n}\vert0\rangle^{\otimes n}$.
\item After applying $n$ Hadamard gates on the first $n$ qubits:
    \begin{displaymath}
        \vert\psi_{1}\rangle=  \frac{1}{\sqrt{2^n}}\sum_{x\in\{0,1\}^n}
            \vert x \rangle \vert0\rangle^{\otimes n}.
    \end{displaymath}
\item Applying the quantum oracle $f$ yields:
    \begin{displaymath}
        \vert\psi_{2}\rangle=  \frac{1}{\sqrt{2^n}}\sum_{x\in\{0,1\}^n}
            \vert x \rangle \vert f(x) \rangle.
    \end{displaymath}
\item We now measure the last $n$ qubits and obtain a certain $f(x_0) \in \{0,1\}^n$, resulting in the (n-qubit) state:
    \begin{displaymath}
        \vert\psi_{3}\rangle=  \frac{1}{\sqrt{2}}( \vert x_0 \rangle + \vert x_0 \oplus a \rangle).
    \end{displaymath}
\item Applying $n$ Hadamard gates on the $n$ qubits yields:
\begin{eqnarray*}
        \vert\psi_{4}\rangle &=&  \frac{1}{2^{(n+1)/2}}\sum_{y\in\{0,1\}^n}
                \big[  (-1)^{x_0 \cdot y} + (-1)^{(x_0 \oplus a )\cdot y} \big] \vert y \rangle\\
                &=&  \frac{1}{2^{(n-1)/2}}\sum_{a \cdot y = 0}
                (-1)^{x_0 \cdot y} \vert y \rangle.
\end{eqnarray*}
\item Measure $\vert\psi_{4}\rangle$ to get a $y$ such that $a \cdot y = 0$.
\end{enumerate}

Repeating steps 1-6 a polynomial number of times will, with high
probability, result in $n$ linearly-independent values $\{y_1,y_2,
\ldots , y_n\}$ such that $y_i \cdot a = 0$, which determines a.

\begin{prop}
Any instance of the Simon problem generates entanglement in Simon's algorithm.
\end{prop}
\textbf{Proof.} To see this, it suffices to examine
$\vert\psi_{2}\rangle$. We show that if it is separable, $f(x)$
must be constant, and therefore it is a trivial case of the
problem. Observe that in $\vert\psi_{2}\rangle$, the first $n$
qubits assume any of their possible values exactly once in the
sum. In order to achieve this state from a tensor product of $2n$
qubits, each of the first $n$ qubits must be of the form $\alpha
\vert 0 \rangle + \beta \vert 1 \rangle$ with $\alpha,\beta \neq
0$. It can therefore be written as:
\[
\ket{\psi_2} = \big( \sum_{x \in \{0,1\}^n}\gamma_x \vert x
\rangle \big) \otimes \big( \sum_{y \in \{0,1\}^n}\delta_y \vert y
\rangle \big).
\]
However, the state of the first $n$ qubits already contributes
$2^n$ \emph{different} elements to the final sum, no matter in
what state the last $n$ qubits are. This means that in order to
have exactly $2^n$ elements in the final sum, the last $n$ qubits
must be in a single computational basis element. Thus, the value
of $f(x)$ is the same for any input, i.e. it is constant.
\hfill$\square$

We thus conclude that the usage of the Simon algorithm for any
subproblem of the Simon problem requires entanglement. However, an
interesting open question in this context, is whether there is a
restricted version of the problem solvable by some \emph{other}
quantum algorithm without entanglement, and achieves an advantage
over the classical case.

\section{Grover's Search Algorithm}
\label{sec:Grover}

Grover presented an algorithm~\cite{Grover96}, which for a binary
function $f:\{0,1\}^n \rightarrow \{0,1\}$, finds an $x$ such that
$f(x)=1$ with only $O(\sqrt{2^n})$ oracle calls.

The first step of the algorithm is identical to the Deutsch-Jozsa
subroutine:
\[
 \frac1{\sqrt N}\sum \ket x \rightarrow \frac1{\sqrt N}\sum (-1)^{f(x)} \ket x.
\]
We require that the resulting state ($\ket{\psi_2}$) will be
separable. In the following we show that this may be true only for
trivial instances of the search problems and therefore we do not
need to follow additional steps of the algorithm. From the
separability of $\ket{\psi_2}$ it follows that if the $n-1$ most
significant bits are measured, the state of the least significant
bit is independent of the measurement outcome. Writing $x=w0$ or
$x=w1$ depending on the value of the
    LSB, $\ket{\psi_2}=1/{\sqrt N}\sum_w \left[(-1)^{f(w0)}\ket {w0}+(-1)^{f(w1)}
    \ket{w1}\right]$ and performing the measurement, we
    are left with the state
    \[
        \frac{(-1)^{f(w0)} \ket0   + (-1)^{f(w1)} \ket1}{\sqrt2}.
    \]
    Up to a global phase, the state is
    \[
        \frac{\ket0 + (-1)^{f(w0)\oplus f(w1)} \ket1}{\sqrt2}
    \]
    and has to be independent of the measured $w$.
    That means that $\forall x: (-1)^{f(x)\oplus
    f(x\oplus 1)}=k$, or more conveniently put as $\forall x: f(x)\oplus f(x\oplus
    1)=y\cdot e_1$. Similarly, entanglement should also be avoided
    when regarding the $j$th qubit for $1\leq j\le n$, which leads
    to
    \[\forall x,j: f(x)\oplus f(x\oplus e_j)=y\cdot e_j.\]
    This means that it is true for a couple of $j$'s combined
\[f(x\oplus e_j)+f(x\oplus e_j\oplus e_i)=y\cdot e_i \Longrightarrow
  f(x)\oplus f(x\oplus e_j \oplus e_i)=y\cdot e_i\oplus y\cdot e_j\]
In the same manner, we can see that for every $J$ it holds that
\[\forall x: f(x)\oplus f(x\oplus J)=y\cdot J\]
even for $x=0$. From this follows that functions that do not
generate entanglement must satisfy
\[f(J)=y\cdot J \oplus f(0),\]
which is the exact definition of the set $F_{DJ}^{\otimes}$. These
functions correspond to trivial search problems where none, all,
or half of the elements are to be found. Hence, any interesting
instance of Grover's search problem would generate entanglement in
Grover's algorithm.

One may confirm that this is true even for $n=2$ bits, regardless
of the fact that only a single oracle call is required in that
case, and unlike what is commented by~\cite{BraunsteinPati02}.
With two bits,
$\ket{\psi_1}=\frac12[\ket{00}+\ket{01}+\ket{10}+\ket{11}]$. If
only one of $\{00,01,10,11\}$ is ``marked'', then $\ket{\psi_2}$
has 3 positive coefficients and a single negative coefficient.
This means that
$\ket{\psi_2}=A\ket{00}+B\ket{01}+C\ket{10}+D\ket{11}$ is
entangled, as it cannot satisfy the separability constraint
$AD=BC$.

\section{Summary and Open Questions}\label{sec:summary}

We investigated the advantage of quantum algorithms without
entanglement over classical algorithms, and showed a maximal
entanglement-free subproblem of the Deutsch-Jozsa problem, which
yields an $O(1)$ to $O(n)$ quantum advantage over the best exact
classical algorithm. Due to this ban on entanglement, the
exponential advantage of exact-quantum versus exact-classical is
lost. For the Simon problem we showed that any non-trivial
subproblem requires entanglement during the computation. Using a
somewhat different approach, we showed that this also holds for
Grover's algorithm.

Further research on the role of entanglement in quantum
information processing may illuminate some of the following
questions: is there a restricted form of the Simon problem (or
more generally of the hidden subgroup problem~\cite{Nielsen00}),
and a corresponding quantum algorithm that presents a quantum
advantage without entanglement? Is there a subproblem larger than
${DJ}^{\otimes}$ and a corresponding algorithm (not the DJ
algorithm) that solves it without entanglement, and yet has an
advantage over any classical algorithm? Can there be a
non-negligible advantage of QCWE over classical computation when
separable mixed states are used? Can it be proved that exponential
advantage of exact QCWE over classical-exact computation is
impossible in oracle-based settings? Note that this is not known
yet, even for the Deutsch-Jozsa problem, since there may be some
other quantum procedure for which the QCWE advantage holds, and
the subset is larger than $F_{DJ}^{\otimes}$.

This work is supported 
in parts by the Israel MOD Research and Technology Unit, by the Institute for
Future Defense Research, and by the Israel Science Foundation --- FIRST
(grant\#4088103).


\section*{References}


\begin{thebibliography}{10}

\bibitem{Shor97}
P.~W. Shor, ``Polynomial-time algorithms for prime factorization and discrete
  logarithms on a quantum computer,'' {\em SIAM Journal on Computing}~{\bf
  26}(5), pp.~1484--1509, 1997.

\bibitem{Grover96}
L.~K. Grover, ``A fast quantum mechanical algorithm for database search,'' in
  {\em Proceedings of the Twenty-Eighth Annual {ACM} Symposium on Theory of
  Computing},  pp.~212--219, ACM Press, (New York), 22--24 May~1996.

\bibitem{BB84}
C.~H. Bennett and G.~Brassard, ``Quantum cryptography: Public key distribution
  and coin tossing,'' in {\em Proceedings of IEEE International Conference on
  Computers, Systems and Signal Processing, Bangalore, India},  pp.~175--179,
  Dec.~1984.

\bibitem{Teleport}
C.~H. Bennett, G.~Brassard, C.~Cr\'epeau, R.~Jozsa, A.~Peres, and W.~K.
  Wootters, ``Teleporting an unknown quantum state via dual classical and
  {Einstein-Podolsky-Rosen} channels,'' {\em Phys. Rev. Lett.}~{\bf 70},
  pp.~1895--1899, Mar.~1993.

\bibitem{DJ92}
D.~Deutsch and R.~Jozsa, ``Rapid solution of problems by quantum computation,''
  {\em Proceedings of the Royal Society of London, Series~A}~{\bf A439},
  pp.~553--558, 1992.

\bibitem{Simon94}
D.~R. Simon, ``On the power of quantum computation,'' {\em SIAM Journal on
  Computing}~{\bf 26}(5), pp.~1474--1483, 1997.

\bibitem{Superdense}
C.~H. Bennett and S.~J. Wiesner, ``Communication via one- and two-particle
  operators on {Einstein-Podolsky-Rosen} states,'' {\em Phys. Rev. Lett.}~{\bf
  69}(20), pp.~2881--2884, 1992.

\bibitem{Shor95}
P.~W. Shor, ``Scheme for reducing decoherence in quantum computer memory,''
  {\em Phys. Rev. A}~{\bf 52}, pp.~R2493--R2496, Oct.~1995.

\bibitem{Ekert91}
A.~K. Ekert, ``Quantum cryptography based on bell's theorem,'' {\em Phys. Rev.
  A}~{\bf 67}, pp.~661--663, 1991.

\bibitem{BBM92}
C.~H. Bennett, G.~Brassard, and N.~D. Mermin, ``Quantum cryptography without
  bell's theorem,'' {\em Phys. Rev. Lett.}~{\bf 68}, pp.~557--559, 1992.

\bibitem{BHM96}
E.~Biham, B.~Huttner, and T.~Mor, ``Quantum cryptographic networks based on
  quantum memories,'' {\em Phys. Rev. A}~{\bf 54}, pp.~2651--2658, 1996.

\bibitem{MHorodecki01}
M.~Horodecki, ``Entanglement measures,'' {\em Quantum Information and
  Computation}~{\bf 1}(1), pp.~3--26, 2001.

\bibitem{Jozsa98book}
R.~Jozsa, {\em Entanglement and Quantum Computation}.
\newblock Oxford University Press, January~1998.

\bibitem{JL02}
R.~Jozsa and N.~Linden, ``On the role of entanglement in quantum computation
  speed-up,'' {\em Proceeding of the Royal Society of {London} series
  {A}---Mathematical Physical and Engineering Sciences}~{\bf 459},
  pp.~2011--2032, Aug.~2003.

\bibitem{Collins98}
D.~Collins, K.~W. Kim, and W.~C. Holton, ``{Deutsch-Jozsa} algorithm as a test
  of quantum computation,'' {\em Phys. Rev. A}~{\bf 58}, p.~1633(R),
  Sept.~1998.

\bibitem{BraunsteinPati02}
S.~L. Braunstein and A.~K. Pati, ``Speed-up and entanglement in quantum
  searching,'' {\em Quantum Information and Computation}~{\bf 2}, pp.~399--409,
  2002.
\newblock Also in quant-ph/0008018.

\bibitem{Lloyd00}
S.~Lloyd, ``Quantum search without entanglement,'' {\em Phys. Rev. A}~{\bf 61},
  p.~010301(R), Dec.~1999.

\bibitem{BBKM02}
E.~Biham, G.~Brassard, D.~Kenigsberg, and T.~Mor, ``Quantum computing without
  entanglement,'' {\em Theoretical Computer Science}~{\bf 320}, pp.~13--33,
  2004.

\bibitem{Meyer00soph}
D.~A. Meyer, ``Sophisticated quantum search without entanglement,'' {\em Phys.
  Rev. Lett.}~{\bf 85}, pp.~2014--2017, 2000.

\bibitem{BV97}
E.~Bernstein and U.~Vazirani, ``Quantum complexity theory$^\dag$,'' {\em SIAM
  Journal on Computing}~{\bf 26}(5), pp.~1411--1473, 1997.

\bibitem{TerhalSmolin98}
B.~Terhal and J.~A. Smolin, ``Single quantum querying of a database,'' {\em
  Phys. Rev. A}~{\bf 58}, pp.~1822--1826, 1998.

\bibitem{Nielsen00}
M.~A. Nielsen and I.~L. Chuang, {\em Quantum Computation and Quantum
  Information}, Cambridge University Press, 2000.

\end{thebibliography}
\end{document}